\documentclass[iop]{emulateapj}

\shorttitle{The H$_2$ emission and the stellar kinematics in the nucleus of M104}
\shortauthors{Menezes \& Steiner}

\begin{document}

\title{The molecular H$_2$ emission and the stellar kinematics in the nuclear region of the Sombrero galaxy}

\author{R. B. Menezes and J. E. Steiner}

\affil{Instituto de Astronomia Geof\'isica e Ci\^encias Atmosf\'ericas, Universidade de S\~ao Paulo, Rua do Mat\~ao 1226, Cidade Universit\'aria, S\~ao Paulo, SP CEP 05508-090, Brazil;}
\email{robertobm@astro.iag.usp.br}

\begin{abstract}

We analyze the molecular H$_2$ emission and the stellar kinematics in a data cube of the nuclear region of M104, the Sombrero galaxy, obtained with NIFS on the Gemini-north telescope. After a careful subtraction of the stellar continuum, the only emission line we detected in the data cube was H$_2 \lambda 21218$. An analysis of this emission revealed the existence of a rotating molecular torus/disk, aproximately co-planar with a dusty structure detected by us in a previous work. We interpret these two structures as being associated with the same obscuring torus/disk. The kinematic maps provided by the Penalized Pixel Fitting method revealed that the stellar kinematics in the nuclear region of M104 appears to be the result of the superposition of a ``cold'' rotating disk and a ``hot'' bulge. Using a model of a thin eccentric disk, we reproduced the main properties of the maps of the stellar radial velocity and of the stellar velocity dispersion, specially within a distance of $0\arcsec\!\!.2$ from the kinematic axis (in regions at larger distances, the limitations of a model of a thin rotating disk become more visible). The general behavior of the $h_3$ map, which is significantly noisier than the other maps, was also reproduced by our model (although the discrepancies, in this case, are considerably higher). With our model, we obtained a mass of \textbf{$(9.0 \pm 2.0) \times 10^8 M_{\sun}$} for the supermassive black hole of M104, which is compatible, at $1\sigma$ or $2\sigma$ levels, with the values obtained by previous studies.

\end{abstract}


\section{Introduction}

Due to its proximity and brightness, M104 (NGC 4594), the Sombrero galaxy, has been observed in many spectral bands. It is an SA(s)a galaxy with a very inclined (nearly edge-on) disk, at a distance of 9.2 Mpc. Its optical spectrum shows a number of low ionization emission lines and, based on this property, the nucleus of this object has been classified as a Low Ionization Nuclear Emission-line Region (LINER - Heckman 1980).

Radio observations have shown that M104 has a compact nuclear source ($\le$ 1pc). \citet{hum84}, using observations of the continuum at 2.6 and 20 cm obtained with the \textit{Very Large Array} (\textit{VLA}), concluded that the central source has a flat spectrum and \citet{baj88}, using data at 1.49 GHz obtained with the \textit{VLA}, detected a variable continuum in that source.

An X-ray compact nuclear source was also observed. \citet{pel02} analyzed \textit{Beppo SAX} and \textit{Chandra} data and verified that the hard X-ray spectrum of the central source is well described by a power law. The authors obtained a luminosity of $L_x = 1.2 - 2.3 \times 10^{40}$ erg s$^{-1}$, in the spectral range 2 - 10 keV, for that source. All these radio and X-ray properties are compatible with the existence of an active galactic nucleus (AGN) in M104.

\citet{nic98}, using data obtained with the \textit{Hubble Space Telescope} (\textit{HST}) and the \textit{ASCA}, concluded that the model of photoionizaton by a low luminosity AGN \citep{fer83,hal83} is the most adequate to explain the emission-line spectrum of this galaxy. In \citet{men13}, we analyzed a data cube of the nuclear region of M104, obtained with the Integral Field Unity (IFU) of the Gemini Multi-object Spectrograph of the Gemini-south telescope, and reported the discovery of collimation and scattering of the AGN emission. We also detected a broad component of the H$\alpha$ emission line, which is another evidence for the existence of an AGN in M104. Such broad component was previously reported by \citet{kor96}, using \textit{HST} data.

\citet{ben06}, using infrared data obtained with the \textit{Spitzer} and with the \textit{James Clerk Maxwell Telescope} of the \textit{Submillimetre Common-User Bolometer Array}, verified that the spectral energy distribution of the nucleus of this galaxy reveals the existence of hot dust. Besides that, the authors also concluded that the low flux at 160 $\mu$m seems to indicate that the LINER emission is a consequence of the lack of cold gas to feed a possible more intense Seyfert activity.

The mass of the supermassive black hole (SBH) in the nucleus of M104 was one of the first SBH masses to be determined. \citet{kor88} analyzed data obtained with the \textit{Canada-France-Hawaii Telescope} (\textit{CFHT}) and found evidence for a SBH with a mass of $10^9 M_{\sun}$. This same value was also obtained by \citet{ems94}, using axisymmetric Jeans models (a generalization of the Jeans formalism is described by Cappellari 2008) and additional \textit{CFHT} data, and by \citet{kor96}, using \textit{HST} and higher resolution \textit{CFHT} data. \citet{mag98}, using also axisymmetric Jeans models, \textit{HST} photometry, and kinematics obtained from ground based observations, determined a mass of $(6.47^{+0.08}_{-0.19}) \times 10^8 M_{\sun}$ for the SBH. Finally, \citet{jar11}, using high-resolution \textit{HST} spectra, long-slit spectra obtained with GNIRS of the Gemini telescope, and also integral-field kinematics obtained with SAURON, applied the Schwarzschild method of superposition of orbits \citep{sch79} and obtained a SBH mass of $(6.6 \pm 0.4) \times 10^8 M_{\sun}$.

In this paper, we present an analysis of a data cube of the nuclear region of M104, in the \textit{K} band, obtained with the Near-infrared Integral Field Spectrograph (NIFS) on the Gemini-north telescope. The use of the adaptive optics (AO) during the observation resulted in a high spatial resolution for the data cube, which also has a combination of high signal-to-noise and spectral resolution. In order to improve the quality of the results, we treated and analyzed the data cube with a variety of techniques. Our goals are: to analyze the emission-line spectrum of the nuclear region of M104 in the \textit{K} band and to analyze the stellar kinematics around the nucleus, with the additional purpose of obtaining an estimate for the mass of the nuclear SBH. 

This paper is organized in the following way: in Section 2, we describe the observations, the data reduction, and the data treatment. In Section 3, we analyze the emission-line spectrum of the data cube. In Section 4, we analyze the stellar kinematics and apply a dynamical modeling, in order to reproduce the observed kinematic maps and to obtain an estimate for the mass of the central SBH. In Section 5, we discuss our results and make comparisons with previous studies. Finally, we present our conclusions in Section 6.

\section{Observations, reduction and data treatment}

\begin{figure*}
\epsscale{0.92}
\plotone{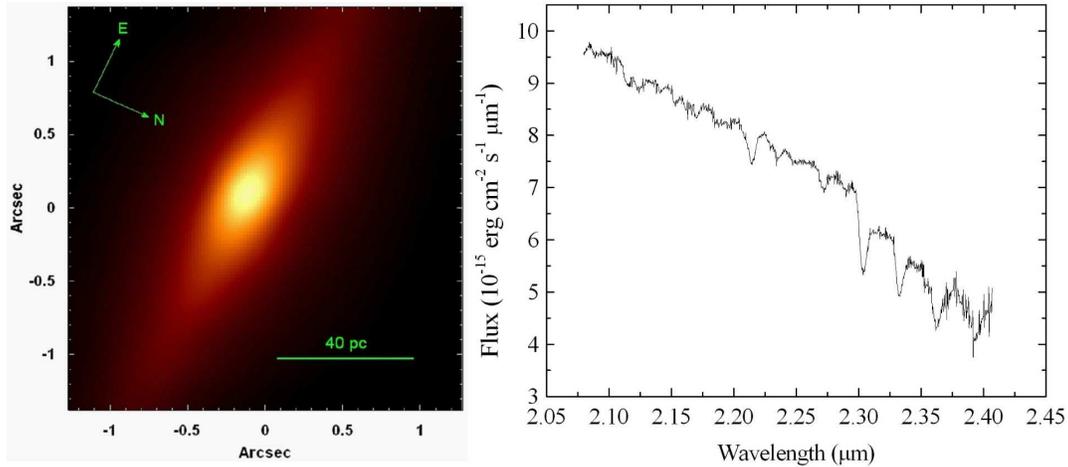}
\caption{Left: image of the treated data cube of M104 collapsed along the spectral axis. Right: average spectrum of the treated data cube of M104.\label{fig1}}
\end{figure*}

The observations of the nuclear region of M104 were made on 2011 April 26, (Programme: GN-2011A-Q-83) with NIFS on the Gemini-north telescope. This instrument has a field of view (FOV) of $3\arcsec \times 3\arcsec$ and works together with the AO module ALTAIR, which allows a maximum spatial resolution of $0\arcsec\!\!.1$. The final products from observations made with this instrument are data cubes, containing two spatial dimensions and one spectral dimension. Five 640 second exposures of the nuclear region of M104 were made, in the \textit{K} band, with a central wavelength of 2.2 $\mu$m. The final spectra had a coverage of 2.01 - 2.42 $\mu$m and a resolution of $R \sim 5155$.

The following calibration images were obtained during the observations: flat-field, dark flat-field (used for dark current subtraction), Ronchi-flat (used to compute spatial distortions and to perform a spatial rectification), arc lamp and sky field. The A0V standard star HIP 56901 was also observed to be used for the telluric absorption removal and for the flux calibration. The data reduction was made with the Gemini IRAF package and included: determination of the trim, sky subtraction, bad pixel correction, flat-field correction, spatial rectification, wavelength calibration, telluric absorption removal, flux calibration (considering that HIP 56901 has a blackbody spectrum), and data cube construction. Five data cubes of M104, with spatial pixels (spaxels) of $0\arcsec\!\!.05 \times 0\arcsec\!\!.05$, were obtained.

We applied a correction of the differential atmospheric refraction (DAR) to all the obtained data cubes, using the practical approach described in \citet{men14}. Then, in order to combine the five corrected data cubes into one, we divided them in two groups. Numbering the data cubes as 1, 2, 3, 4, and 5 (following the order of the observations), we applied the following division: \\
\\
Group 1: 1, 2, and 5 \\
Group 2: 1, 3, and 4.\\
\\
A median of each group was calculated and, after that, an average of the two obtained data cubes was calculated, resulting in the combined data cube.

\begin{figure*}
\epsscale{0.8}
\plotone{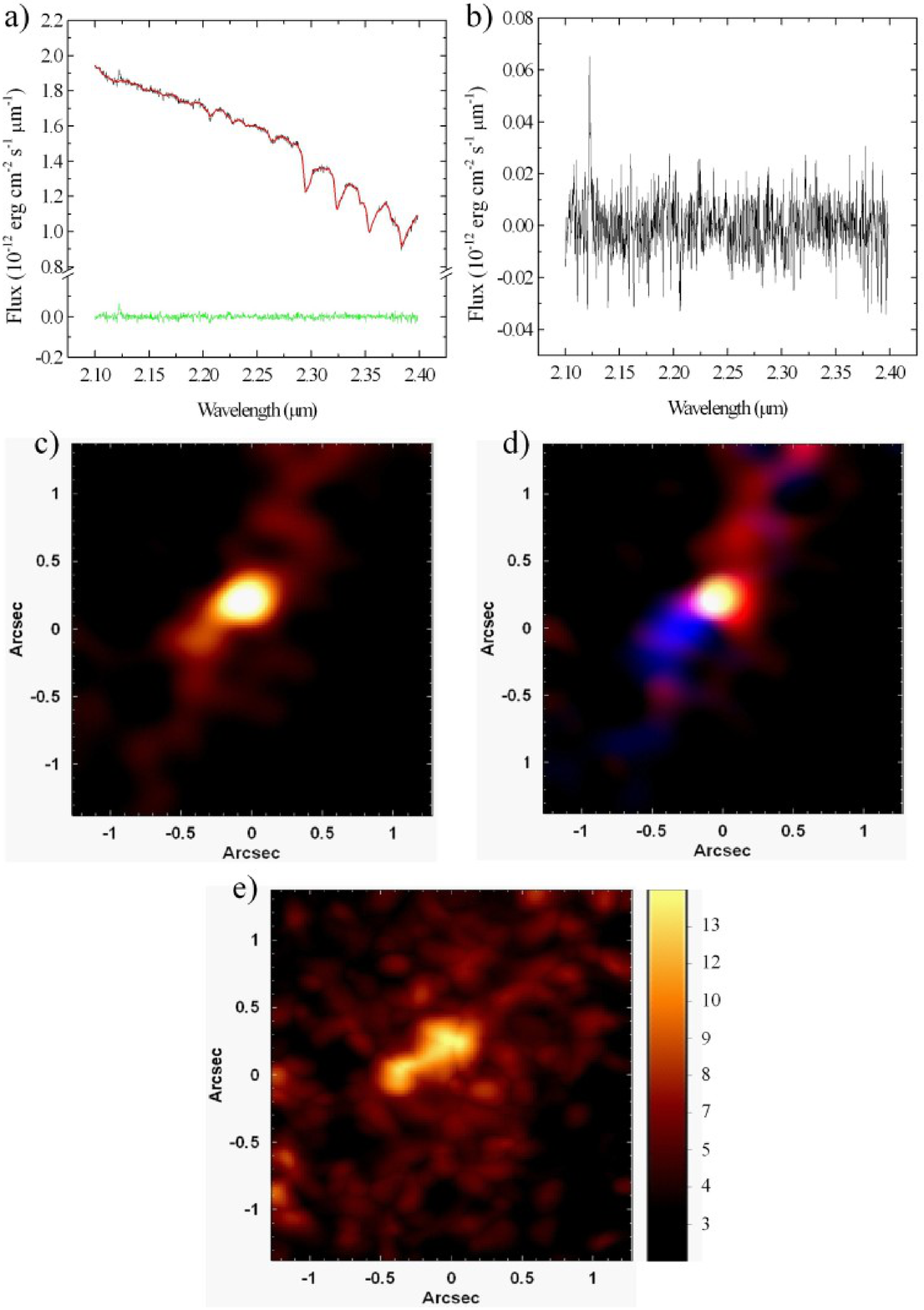}
\caption{(a) Spectrum from a circular area, with a radius of $0\arcsec\!\!.084$, centered on the nucleus of the stellar bulge in the data cube of M104. The fit provided by the pPXF is shown in red and the fit residuals are shown in green. (b) Magnification of the same fit residuals shown at left. (c) Image of the H$_2 \lambda 21218$ emission line of the data cube of M104, after the starlight subtraction. (d) RGB composite image of the H$_2 \lambda 21218$ emission line. The colors blue, green, and red correspond to the velocity ranges $-438 \le V_r \le -170$, $-156 \le V_r \le 156$, and $170 \le V_r \le 438$ km s$^{-1}$, respectively. (e) Amplitude/noise ratio map of the H$_2 \lambda 21218$ emission line.\label{fig2}}
\end{figure*}

In order to improve the visualization of the contours of the structures, we performed a spatial re-sampling to all the images of the combined data cube (conserving the surface flux), to obtain spaxels of $0\arcsec\!\!.021$. The improvement of the appearance of the images in the data cube is not only reason to apply the spatial re-sampling. We verified that the final spatial resolution is higher when the Richardson-Lucy deconvolution (see more details below) is applied to a spatially re-sampled data cube. After that, we applied a Butterworth spatial filtering to the re-sampled images of the data cube, in order to remove high spatial frequency noise. As explained in \citet{men14}, we used a Butterworth filter corresponding to the product between an elliptic filter and a rectangular one. The order of each one of these filters was $n = 2$ and their cutoff frequency was 0.19 Ny (Nyquist frequency).

The data cube of M104 revealed the existence of an ``instrumental fingerprint'', in the form of large vertical stripes across the images. This feature also has a specific low-frequency spectral signature. The cause of the instrumental fingerprint on NIFS data cubes is still unknown. As explained in \citet{men14}, NIFS reads out through four amplifiers in 512 pixel blocks in the spectral direction. Therefore, a slight bias level difference between these amplifiers could be an explanation for this fingerprint. We used the Principal Component Analysis Tomography technique \citep{ste09} to remove the fingerprint from the data cube of M104, following the prescription of \citet{men14,men15}. 

The last step of our data treatment procedure was the Richardson-Lucy deconvolution \citep{ric72,luc74}, which was applied to all the images of the data cube, in order to improve the spatial resolution of the observation. The point-spread function (PSF) used for this process was an image of the data cube of the star HIP 56901. The use of an image of the standard star data cube was necessary, because we were not able to determine any other reliable estimate for the PSF. This strategy requires much caution, as there can be differences between the effects of the AO applied to the science data cube and to the standard star data cube. Nevertheless, our tests revealed that, if such differences existed, in this case, they were of second order and, as a consequence, the Richardson-Lucy deconvolution not only did not compromise the data cube of M104, but actually improved significantly its spatial resolution. A comparison with an \textit{I}-band image of M104 obtained with the Wide-field Planetary Camera 2 (WFPC2) of the \textit{HST} revealed that the FWHM of the PSF of the data cube was $\sim 0\arcsec\!\!.3$, before the Richardson-Lucy deconvolution, and $\sim 0\arcsec\!\!.15$, after the Richardson-Lucy deconvolution. For more details about the data treatment procedure we performed, see \citet{men14,men15}. An image of the treated data cube of M104 collapsed along the spectral axis, together with its average spectrum, is shown in Figure~\ref{fig1}.

The main feature in the image of the collapsed data cube of M104 is an elongated structure with a position angle (PA) of $63\degr \pm 4\degr$. This is the inner stellar disk of this galaxy, which was analyzed in previous studies \citep{bur86,ems94, ems96}.

\section{Analysis of the emission line spectrum}

One of our interests in the data cube of M104 is the analysis of the emission lines in the nuclear region of this galaxy. However, this analysis requires an accurate starlight subtraction. In order to perform this procedure, we applied the Penalized Pixel Fitting (pPXF) method \citep{cap04} to all the spectra of the data cube. This procedure uses a combination of template spectra from a given base convolved with a Gauss-Hermite expansion, to obtain the final synthetic spectrum. In this case, we used a base of stellar spectra observed with NIFS, described in detail in \citet{win09}. Since this base has the same spectral resolution of our data cube, it is very appropriate to be used to fit the absorption lines in the data cube. On the other hand, this base does not provide a good fit for the stellar continuum. Therefore, the algorithm we used to apply the pPXF added a fourth degree Legendre polynomial to the combination of the template spectra from the base, in order to fit the stellar continuum. 

Before applying the pPXF to the data cube of M104, we shifted the spectra to the rest-frame, using $z = 0.003416$ (Nasa Extragalactic Database - NED), and sampled them with $\Delta\lambda = 1\AA$ per pixel. All these steps were performed using scripts written in {\sc IDL}. The parameters provided by the pPXF are: the stellar radial velocity ($V_*$), the stellar velocity dispersion ($\sigma_*$), and the Gauss-Hermite coefficients $h_3$ and $h_4$. The Gauss-Hermite coefficient $h_3$ reveals asymmetries in the stellar absorption lines while the Gauss-Hermite coefficient $h_4$ indicates deviations from a Gaussian profile in these lines \citep{ger93,van93}. A synthetic stellar spectrum is also provided by the pPXF method. The starlight subtraction we required for a detailed analysis of the emission lines in the data cube of M104 was performed by subtracting the obtained synthetic stellar spectra from the observed ones. Figure~\ref{fig2} shows an example of the fit (together with the fit residuals) obtained with the pPXF for a spectrum from a circular area, with a radius of $0\arcsec\!\!.084$, centered on the nucleus of the stellar bulge.

A simple analysis of Figure~\ref{fig2} reveals that the fit reproduced, with good precision, the continuum and the absorption lines. The fit residuals show only one significant emission feature at 2.1218 $\mu$m, which corresponds to the H$_2 \lambda 21218$ emission line. We believe that other H$_2$ lines, like H$_2 \lambda 22234$ or H$_2 \lambda 22478$, were not detected in this case because they are immersed in the noise of the spectrum. The H$_2 \lambda 21218$ emission line was detected here probably because it is usually the most intense H$_2$ line in this spectral region. We extracted the total spectrum of the data cube, after the starlight subtraction, and then we calculated the integrated flux of the H$_2 \lambda 21218$ line, obtaining a value of $F_{H2} = (6.8 \pm 0.7) \times 10^{-15}$ erg cm$^{-2}$ s$^{-1}$. Since we observed only the H$_2 \lambda 21218$ emission line in the data cube of M104, we could not perform a reliable analysis involving the ratios of H$_2$ lines, in order to determine if this molecular emission is due to thermal or non-thermal processes in this galaxy \citep{mou94}.  

In order to analyze the morphology of the H$_2 \lambda 21218$ emitting regions, we constructed an image of this emission line. We also constructed an RGB composite image (based on the radial velocity values) for this line. The results are shown in Figure~\ref{fig2}. We can see that most of the H$_2 \lambda 21218$ emission comes from a structure with PA = $65\degr \pm 9\degr$, which is compatible (at 1$\sigma$ level) with the PA of the inner stellar disk. This, together with the observed morphologies, suggest that the molecular gas is disposed along the inner stellar disk. The center of this molecular gas distribution is located at a projected distance of $\sim 0\arcsec\!\!.12$ (toward east) from the stellar bulge center. Figure~\ref{fig2} also shows an amplitude/noise (A/N) map of the H$_2 \lambda 21218$ emission line. We can see that the A/N values are in the range of 10 - 14 near the stellar bulge center and decrease significantly at farther areas. They are smaller than 3 at the most peripheral regions. In order to construct a velocity map for the H$_2$ molecular gas, we fitted Gaussian functions to the H$_2 \lambda 21218$ emission line. However, with these low A/N ratios, the Gaussian fits did not provide reliable values for the radial velocity. We performed a spatial binning of the data cube but, in order to obtain sufficiently high A/N ratios for the H$_2 \lambda 21218$ emission line, the spatial resolution and the number of resolution elements were severely reduced. Considering all of that, we decided to not include the velocity map of the H$_2 \lambda 21218$ emission line in this paper.

Although we could not construct a reliable kinematic map of the H$_2 \lambda 21218$ emission line, the RGB composite image in Figure~\ref{fig2} indicates that the kinematics of the molecular gas seems to be compatible with a rotation around the nucleus. Therefore, we conclude that the obtained results indicate the existence of a rotating torus/disk structure, approximately co-planar with the inner stellar disk around the nucleus of M104.  

\begin{figure*}
\epsscale{0.9}
\plotone{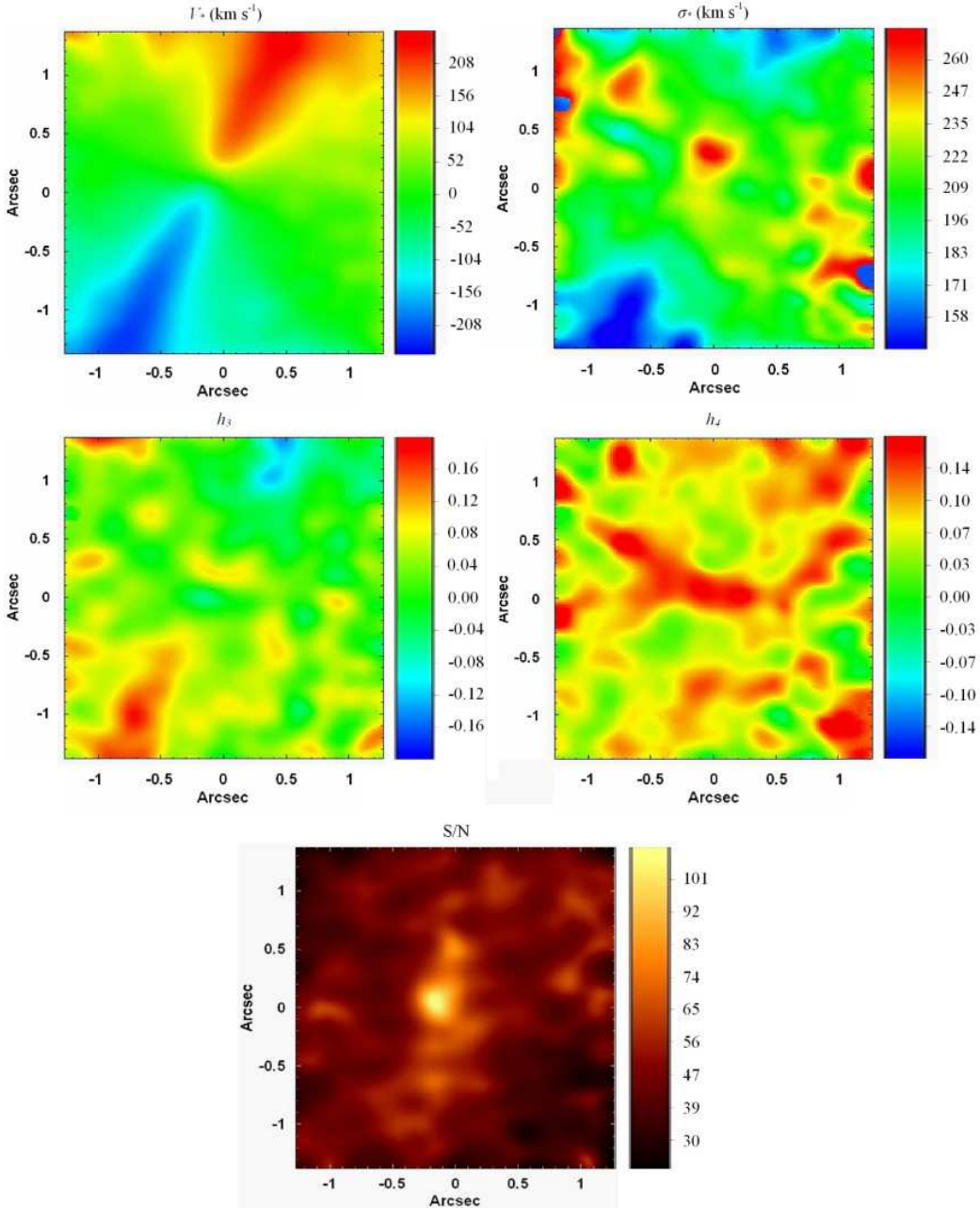}
\caption{Kinematic maps obtained with the pPXF method applied to the data cube of M104. An S/N ratio map for the wavelength range of 2.2730 - 2.2777 $\mu$m is shown at the bottom.\label{fig3}}
\end{figure*}

\section{Analysis of the stellar kinematics}

Our second interest in the data cube of M104 is the analysis of the stellar kinematics around the nucleus. As mentioned before, the pPXF method provides the values of $V_*$, $\sigma_*$, $h_3$, and $h_4$. Therefore, since this method was applied to each spectrum of the data cube of M104, we constructed maps of these parameters, which are shown in Figure~\ref{fig3}. In addition, in order to evaluate if the values obtained for the kinematic parameters are reliable, we constructed an signal-to-noise ratio (S/N) map (for the wavelength range of 2.2730 - 2.2777 $\mu$m). The result is also shown in Figure~\ref{fig3}. It is important to mention that the $V_*$ map was calibrated in order to ensure that $V_*$ was equal to 0 at the stellar bulge center. In other words, the values of $V_*$ shown in the map in Figure~\ref{fig3} are the stellar radial velocities relative to the stellar bulge center.

\begin{figure*}
\epsscale{1.1}
\plotone{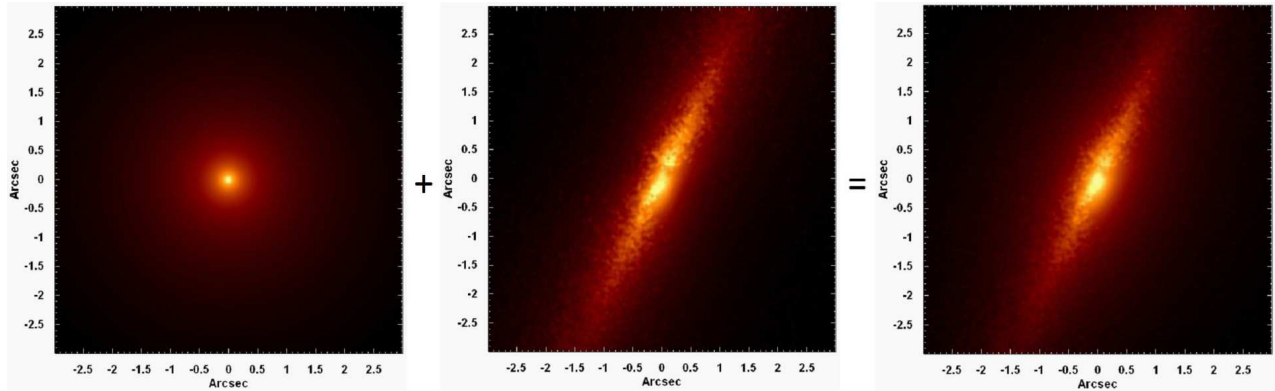}
\caption{Left: image of the symmetric component (with a Sersic index of 3.7) of the nuclear region of M104, obtained with the decomposition of an \textit{HST} image, in the \textit{I} band. Center: image of the asymmetric component of the nuclear region of M104, obtained with the decomposition of the same \textit{HST} image. Right: original \textit{HST} image that was decomposed in a symmetric and in an asymmetric component. In all images, the orientation is the same as in the data cube of M104.\label{fig4}}
\end{figure*}

An analysis of the $V_*$ map in Figure~\ref{fig3} reveals the existence of a rotating stellar disk around the nucleus, with a maximum velocity of $V_*(max) = 256 \pm 7$ km s$^{-1}$ and a minimum velocity of $V_*(min) = -223 \pm 7$ km s$^{-1}$. The $\sigma_*$ map is a little noisier than the $V_*$ map and shows a peak with $\sigma_*(peak) = 272 \pm 9$ km s$^{-1}$, which is located at a projected distance of $\sim 0\arcsec\!\!.21$ (toward east) from the center of the $V_*$ map. The $\sigma_*$ values are smaller along the stellar disk than in regions at larger distances. This is evidence that the superposition of a ``cold'' rotating disk (with lower $\sigma_*$ values) and a ``hot'' bulge (with higher $\sigma_*$ values) is a good description for the stellar kinematics in the nuclear region of M104. Although the $h_3$ map is noisier than the previous two, it is possible to detect a pattern, as the smallest values of $h_3$ (with a minimum of $h_3(min) = -0.18 \pm 0.02$) are located in the region with the largest stellar velocities and the largest values of $h_3$ (with a maximum of $h_3(max) = 0.14 \pm 0.04$) are located in the region with the smallest stellar velocities. Finally, the $h_4$ map does not show clear structures or patterns.

Figure~\ref{fig3} also shows that the S/N values are larger than 30 in most of the FOV of the data cube (they are even larger than 100 near the center of the stellar bulge). Only at the most peripheral areas the S/N values are smaller than 30. This indicates that the features shown in the kinematic maps (like the off-centered $\sigma_*(peak)$, for example) are not artefacts caused by small S/N values. In order to evaluate if some of these kinematic features were introduced by the Richardson-Lucy deconvolution or by the spatial re-sampling of the data cube of M104, we applied the pPXF method to the non-deconvolved data cube, with spaxels of $0\arcsec\!\!.021$, and also to the spatially binned non-deconvolved data cube, with spaxels of $0\arcsec\!\!.1$. The obtained maps had the same characteristics of the ones shown in Figure~\ref{fig3}. Therefore, we conclude that the Richardson-Lucy deconvolution and the spatial re-sampling did not affect the measurement of the stellar kinematics. Finally, we also extracted the kinematics of the data cube of M104 using a Gaussian line of sight velocity distribution. Again, the main features of the $V_*$ and $\sigma_*$ maps (including the off-centered $\sigma_*(peak)$) were compatible with the ones in Figure~\ref{fig3}.

The asymmetry in the $V_*$ map, together with the fact the the $\sigma_*(peak)$ is displaced from the center of the $V_*$ map, are evidences for the existence of a rotating excentric stellar disk in the nuclear region of M104. In order to obtain an estimate for the mass of the SBH in this galaxy, we performed a simple dynamical modeling, assuming a thin eccentric stellar disk around the nucleus. For this procedure, first of all, we superposed 126 concentric elliptic orbits, taking as free parameters the argument of the pericenter $\omega$, the mass of the central black hole $M_{\bullet}$, the inclination of the disk $i$, and the eccentricity of the disk $e$. The longitude of the ascending node was measured in the $V_*$ map ($\Omega = 56\degr\!\!.4$); therefore, we kept this parameter fixed in our model. The calculation of the velocities in each one of the orbits also required the knowledge of the stellar mass within each orbit. In order to determine that, we retrieved an image of the nuclear region of M104, obtained with WFPC2 of the \textit{HST}, using the F814W filter. This image was calibrated in order to provide the values of the luminosity in the \textit{I} band. After that, we decomposed the image into a symmetric component (representing the central region of the stellar bulge of M104) and an asymmetric one (representing the rotating stellar disk). We verified that the best result of this decomposition is obtained when we assume a Sersic profile, with a Sersic index of 3.7, to the central region of the stellar bulge. Figure~\ref{fig4} shows the \textit{HST} image, in the F814W filter, of the nuclear region of M104, together with the images obtained, after the decomposition, for the symmetric and asymmetric components. In order to de-project the symmetric component, we applied a Multi-Gaussian Expansion (Cappellari 2002) and, as a result, obtained the values of the luminosity density (in $L_{I\sun}/pc^3$). Considering that, we were able to determine the stellar mass within each elliptical orbit of our model by assuming a mass to light ratio in the \textit{I} band ($M/L_I$), which was taken as another free parameter. Once all the concentric orbits were superposed, the velocity values corresponding to each one of them were calculated and the orbits were projected on the plane of the sky. As a result, we obtained a velocity map. 

\begin{figure*}
\epsscale{1.0}
\plotone{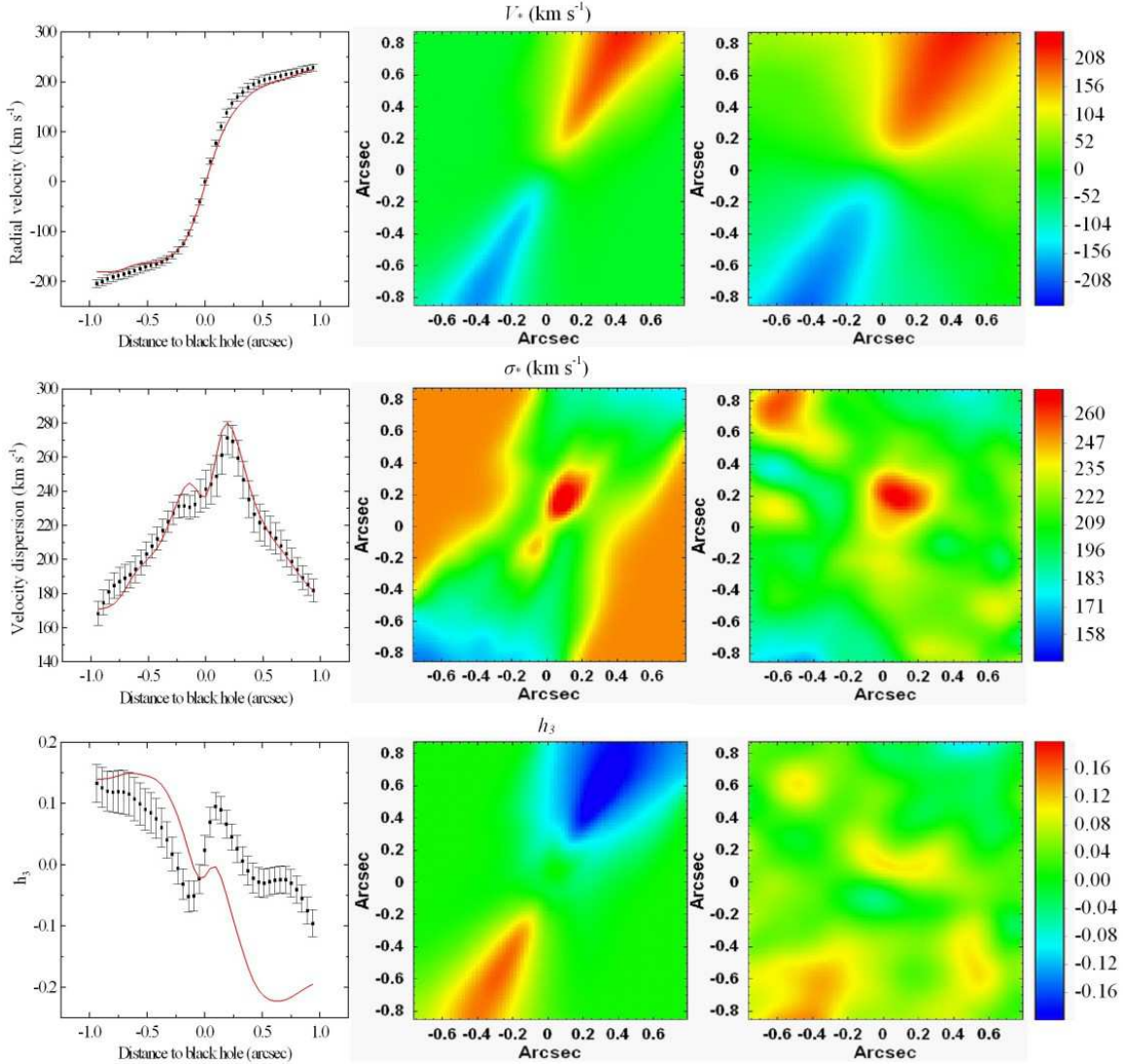}
\caption{Left: curves of $V_*$, $\sigma_*$, and $h_3$, extracted along the kinematic axis. The observed curves are the points with error bars and the simulated curves of the best obtained model are shown in red. Center: simulated maps of $V_*$, $\sigma_*$, and $h_3$ of the best obtained model. Right: observed maps of $V_*$, $\sigma_*$, and $h_3$, with a reduced FOV.\label{fig5}}
\end{figure*}

The next step in our dynamical modeling was the construction of two synthetic data cubes: one representing the rotating stellar disk and the other representing the stellar bulge. In order to do that, we first obtained a representative spectrum for the data cube of M104, which was taken as a linear combination of the spectra of the base used before to apply the pPXF method. The coefficients of this linear combination were provided by the pPXF applied to the average spectrum of the data cube of M104. The representative spectrum was obtained with a spectral range of 2.25 - 2.40 $\mu$m. The synthetic data cube representing the rotating stellar disk was constructed by shifting the representative spectrum using the values of radial velocities given by the synthetic velocity map obtained before and then by placing each shifted spectrum in the position of the corresponding spaxel of the synthetic velocity map. All these spectra were also convolved with an estimate of the stellar velocity dispersion of the rotating disk ($\sigma_d$), which is another free parameter of our model. It is important to emphasize that we assumed a constant stellar velocity dispersion for the rotating stellar disk. The synthetic data cube representing the stellar bulge was constructed by placing the representative spectrum, with a radial velocity equal to 0, in the position of each spaxel. All these spectra were convolved with an estimate of the stellar velocity dispersion of the bulge ($\sigma_b$), which is also a free parameter in our model. In other words, we assumed that the stellar bulge shows a radial velocity equal to 0 and also has a constant velocity dispersion. Using the images of the symmetric and asymmetric components of the nuclear region of M104, obtained from the decomposition of the \textit{HST} image described before, we were able to determine the relative weights between the spaxels of the constructed synthetic data cubes. With these relative weights, the synthetic data cubes representing the rotating stellar disk and the stellar bulge were combined into one. Finally, each image of the combined data cube was convolved with an estimate of the PSF of the observation. In order to obtain the simulated kinematic maps, we applied the pPXF to the synthetic data cube, completing the procedure. In summary, the free parameters of our model are: $\omega$, $M_{\bullet}$, $i$, $e$, $M/L_I$, $\sigma_d$, and $\sigma_b$. Since the $h_4$ map shown in Figure~\ref{fig3} is quite noisy and does not show clear structures, we only tried to reproduce the maps of the values of $V_*$, $\sigma_*$, and $h_3$. Considering that we are more interested in the stellar kinematics near the central black hole and also in order to avoid higher imprecisions in the calculations of the stellar mass, only a region within a radius of $0\arcsec\!\!.85$ was simulated. The free parameters of our model were varied and this process was repeated several times, in order to minimize the value of the total $\chi^2$ ($\chi_T^2$), which is given by
\\
\begin{equation}
\resizebox{.98\hsize}{!}{$\chi^2_T=\frac{(\frac{1}{Min(\chi^2_{V_*})})*\chi^2_{V_*}+(\frac{1}{Min(\chi^2_{\sigma_*})})*\chi^2_{\sigma_*}+(\frac{1}{Min(\chi^2_{h_3})})*\chi^2_{h_3}}{(\frac{1}{Min(\chi^2_{V_*})})+(\frac{1}{Min(\chi^2_{\sigma_*})})+(\frac{1}{Min(\chi^2_{h_3})})}$},
\end{equation}
\\
where $\chi^2_{V_*}$, $\chi^2_{\sigma_*}$, and $\chi^2_{h_3}$ are the $\chi^2$s calculated for the simulated maps of $V_*$, $\sigma_*$, and $h_3$, respectively, and $Min(\chi^2_{V_*})$, $Min(\chi^2_{\sigma_*})$, and $Min(\chi^2_{h_3})$ are the minima values obtained for $\chi^2_{V_*}$, $\chi^2_{\sigma_*}$, and $\chi^2_{h_3}$, respectively. Our tests revealed that, in our dynamical modeling, the weighted mean given by equation (1) is the most effective way to combine the values of $\chi^2_{V_*}$, $\chi^2_{\sigma_*}$, and $\chi^2_{h_3}$ into one.  

\begin{table}
\begin{center}
\caption{Parameters of the Best Model Obtained to Reproduce the Stellar Kinematics of the Data Cube of M104.\label{tbl1}}
\begin{tabular}{cc}
\tableline\tableline
Parameter & Value\\
\tableline
$M_{\bullet}$ & $(9.0 \pm 2.0) \times 10^8 M_{\sun}$\\
$e$ & $0.12 \pm 0.05$\\
$i$ & $80\degr \pm 2\degr$\\
$\omega$ & $5\degr \pm 9\degr$\\
$\sigma_b$ & $(260 \pm 14)$ km s$^{-1}$\\
$\sigma_d$ & $(120 \pm 8)$ km s$^{-1}$\\
$M/L_I$ & $3.0 \pm 0.5$\\ 
\tableline
\end{tabular}
\end{center}
\end{table}

We defined the $\chi^2$ calculated for the simulated map of a kinematic parameter $A_*$ in the following way:
\\
\begin{equation}
\resizebox{.98\hsize}{!}{$\chi^2_{A_*}=\frac{1}{I}\sum_{i=1}^{N_x}\sum_{j=1}^{N_y}\frac{w_{ij}\cdot I_{ij}\cdot \left(A_{ij}(observed)-A_{ij}(simulated)\right)^2}{\sigma_{A_{ij}}^2}$},
\end{equation}
\\
where $A_{ij}(observed)$ and $A_{ij}(simulated)$ are the values of the spaxel $(i,j)$ of the observed and simulated maps of $A_*$, respectively, $N_x$ and $N_y$ are the number of spaxels along the horizontal and vertical axis, respectively, $\sigma_{A_{ij}}$ is the uncertainty of $A_*$ in the spaxel $(i,j)$, $w_{ij}$ is a step function equal to 1 for areas closer than $0\arcsec\!\!.20$ to the kinematical axis and equal to 0 for areas at larger distances, $I_{ij}$ is the integrated flux of the data cube of M104 in the spaxel $(i,j)$, and $I$ is the total integrated flux of the data cube of M104. The values of the uncertainties $\sigma_{A_{ij}}$ of the kinematic parameters analysed in this work ($V_*$, $\sigma_*$, and $h_3$) were obtained via Monte Carlo simulations. The calculations of the values of $\chi^2_{A_*}$ involved the integrated fluxes of the data cube of M104 ($I_{ij}$ and $I$) and also the step function $w_{ij}$, in order to give more weight to the areas closer to the kinematical axis of the data cube. If we had not adopted this strategy, most of the weight in the calculation of $\chi^2_{A_*}$ would have come from the outer parts of the image (with a higher number of spaxels). However, we are more interested in reproducing the kinematic parameters in areas near the kinematic axis. Besides that, since we are applying a model of a thin disk, it is expected that the limitations of our model be more evident in areas farther away from the plane of the disk. Therefore, considering that the rotating disk of M104 is almost edge-on, the limitations of our model would be more significant in regions at larger distances from the kinematic axis, originating higher discrepancies. Based on all of that, we decided to perform the calculation of the values of $\chi^2_{A_*}$ giving more weight to areas near the kinematic axis. There are different ways to give more weight to the central regions of the images during the calculation of the $\chi^2$; however, once again, our tests revealed that the specific calculation given by equation (2) provides the best results in this case.

The simulated maps, and the simulated curves extracted along the kinematic axis, of $V_*$, $\sigma_*$, and $h_3$ of the best obtained model, with $\chi^2_T = 1.519$, are shown in Figure~\ref{fig5}. In the same Figure, we show the corresponding observed maps (with reduced FOV). The kinematic parameters of the best model are shown in Table~\ref{tbl1}. We estimated the uncertainties (1$\sigma$) of the parameters by following the approach described in \citet{der11}:  we plotted an histogram for each parameter of the simulation, considering only cases with $\chi^2_T - \chi^2_{min} < 1$. Then, a Gaussian function was fitted to each histogram. The value obtained for the square deviation of each Gaussian was taken as the uncertainty of the corresponding parameter. 

An analysis of Figure~\ref{fig5} reveals a good agreement between the observed and simulated $V_*$ curves, as all the simulated values are compatible with the observed ones at $1\sigma$ or $2\sigma$ levels. There is also a good agreement between the simulated and observed $V_*$ maps within a distance of $\sim 0\arcsec\!\!.2$ from the kinematic axis. On the other hand, we can see a significant discrepance between the simulated and observed values in areas at distances of $0\arcsec\!\!.2 - 0\arcsec\!\!.4$ from the kinematic axis. As mentioned before, this is probably a consequence of the limitations of a thin disk model. For regions at larger distances, both the simulated and observed maps  show values close to 0. Almost all the observed values of the $\sigma_*$ curve are compatible, at $1\sigma$ level, with the simulated ones. The general morphology of the $\sigma_*$ map, including the position of the peak, is well described the model. However, again, we can see certain discrepancies, most likely related to the limitations of the model,  in areas at distances of $0\arcsec\!\!.2 - 0\arcsec\!\!.4$ from the kinematic axis. For regions at larger distances, the observed $\sigma_*$ map is much more irregular. It is probable that these irregularities represent deviations from the model we used here and that could explain the observed discrepancies. In the case of the values of $h_3$, the situation is somewhat different. Although the general patterns of the observed $h_3$ curve and $h_3$ map are reasonably reproduced by the model, there are many discrepancies that should be discussed. The points with positive values of the simulated and observed $h_3$ curves are compatible at $1\sigma$ or $2\sigma$ levels. However, most of the other points of these curves are not compatible, even at $3\sigma$ level. The observed $h_3$ map is quite noisy and shows several assimetries and irregularities that were not adequately reproduced by our model. Part of these irregularities may be the result of uncertainties of the pPXF method. However, other irregularities may be real and, therefore, represent deviations from our thin disk model. One example is the central features in the $h_3$ map correlated to the $V_*$ map. As explained above, the S/N values in the areas near the stellar bulge center are significantly large. In addition, the $h_3$ maps obtained with the pPXF applied to the non-deconvolved data cube, with spaxels of $0\arcsec\!\!.021$, and to the spatially binned non-deconvolved data cube, with spaxels of $0\arcsec\!\!.1$, also shows these structures correlated to the $V_*$ map. This indicates that these features are not artefacts introduced by the Richardson-Lucy deconvolution or by the spatial re-sampling. Based on this discussion, we conclude that the central structures in the $h_3$ map are real and not a consequence of uncertainties caused by small S/N values or an artefact introduced by the methodologies used during the data treatment. These $h_3$ features correlated to the $V_*$ map are not expected for a rotating disk within a bulge, which suggests the existence of deviations from the model we applied here. The high asymmetry in the h3 map is probably also a deviation from the model, as this behavior was observed in all $h_3$ maps obtained with the applications of the pPXF to the non-deconvolved data cubes (with and without a spatial binning) of M104.

\section{Discussion and comparison with previous studies}

In \citet{men13}, we have detected a dusty dense structure (with PA $= 72\degr\pm14\degr$ and an outer radius of $0\arcsec\!\!.40$, corresponding to $r = 18$ pc) in the nuclear region of M104, which we interpreted as an obscuring torus/disk that collimates the AGN emission of this galaxy. The PA of this torus/disk is compatible, at $1\sigma$ level, with the PA of the rotating H$_2$ torus/disk (PA = $65\degr \pm 9\degr$) observed in this work. This suggests that the structures detected in \citet{men13} and in this work are coplanar and associated with the same obscuring torus/disk. 

The morphology of the $V_*$ map we obtained for the nuclear region of M104 is compatible with the morphology of the map obtained by \citet{ems96}, using the integral-field spectrograph TIGER. The morphologies of our $\sigma_*$ and $h_3$ maps are also similar to what was obtained by these authors, but some discrepancies can be seen. We believe that these discrepancies can be explained by the fact that the spatial resolution of our NIFS data cube ($\sim 0\arcsec\!\!.15$) is significantly higher than the spatial resolution of the TIGER data cube obtained by \citet{ems96} ($\sim 0\arcsec\!\!.9$). Therefore, we are able to detect structures and irregularities that are impossible to be observed in TIGER data. 

The observed rotation curve shown in Figure~\ref{fig5} is consistent with the curves obtained by \citet{ems96} and by \citet{kor96}. These previous works also obtained $\sigma_*$ curves, which are similar to our $\sigma_*$ curve, but some differences can be easily detected. Our $\sigma_*(peak)$ ($= 272 \pm 9$ km s$^{-1}$) is nearly compatible, at $3\sigma$ level, with the $\sigma_*(peak)$ of \citet{ems96} ($= 248$ km s$^{-1}$) and is not compatible with the $\sigma_*(peak)$ obtained by \citet{kor96} ($= 321 \pm 7$ km s$^{-1}$). We believe that these discrepancies are caused by differences in the spatial resolutions of these works. One interesting topic is that the positions of $\sigma_*(peak)$ determined by \citet{ems96} and \citet{kor96} are apparently coincident with the position of the $V_*$ map center. However, in the present work, we observed a projected distance of $\sim 0\arcsec\!\!.21$ between the position of $\sigma_*(peak)$ and the $V_*$ map center. The separation of these points could not be detected by \citet{ems96} due to the spatial resolution of the observation. On the other hand, the non-detection of this behavior of $\sigma_*(peak)$ by \citet{kor96} is more difficult to explain. Three points of the $\sigma_*$ curve of M104 presented by \citet{kor96} were obtained with the Faint Object Spectrograph (FOS) of the \textit{HST} (while the other points were obtained with the Subarcsecond Imaging Spectrograph of the \textit{CFHT}). These three FOS values of $\sigma_*$ were determined from spectra resulting from three observations, using the $0\arcsec\!\!.21$ square aperture, at distances of $0\arcsec\!\!.032$ W, $0\arcsec\!\!.163$ W, and $0\arcsec\!\!.207$ E from the nucleus. All these observations were taken along the east-west direction. Based on this information, we conclude that the spectrum of the region at $0\arcsec\!\!.207$ E from the nucleus could have revealed the off-centered $\sigma_*(peak)$ we observed in the data cube of M104. However, one should note that the $\sigma_*$ curve in \citet{kor96} reveals that the difference between the $\sigma_*$ values provided by the spectra taken at $0\arcsec\!\!.032$ W and at $0\arcsec\!\!.207$ E from the nucleus are $\sim 10$ km s$^{-1}$. On the other hand, the error bars of the point at a radius of $0\arcsec\!\!.032$ W are $\sim 25$ km s$^{-1}$ while the error bars of the point at a radius of $0\arcsec\!\!.207$ E are $\sim 10$ km s$^{-1}$. Therefore, considering the difference between the $\sigma_*$ values in these two points and the error bars, we conclude that the separation between the $\sigma_*(peak)$ and the stellar bulge center we observed in the data cube of M104 is actually not incompatible with the results obtained by \citet{kor96}. There are also significant discrepancies between the $h_3$ curves obtained by us and by \citet{ems96}. However, considering, again, the differences in the spatial resolutions of the corresponding observations, we can say that the observed discrepancies are not surprising. The central features in the $h_3$ map correlated to the $V_*$ map may be indicative of a bar in the central region of M104 \citep{chu04,ian15}. The hypothesis of the existence of a bar in M104 was already discussed by \citet{ems00} and should be taken into account in future studies.

The mass of the SBH of M104 we obtained with a simple model of a thin eccentric disk (\textbf{$9.0 \pm 2.0 \times 10^8 M_{\sun}$}) is compatible, at $1\sigma$ level, with the value of $10^9 M_{\sun}$ estimated by \citet{kor88}, \citet{ems94}, and \citet{kor96}. Our result is also compatible, at $2\sigma$ level, with the masses obtained by \citet{jar11} ($6.6 \pm 0.4 \times 10^8 M_{\sun}$) and by \citet{mag98} ($6.47^{+0.08}_{-0.19} \times 10^8 M_{\sun}$). The value of $M/L_I$ resulting from our model is compatible, at 1$\sigma$ level, with the value obtained by \citet{jar11} ($M/L_I = 3.40 \pm 0.05$). Our model is certainly much more simplistic than other dynamical models, like the Schwarzschild and the Jeans methods, which reproduce in detail the gravitational potential around the SBH. The Schwarzschild method involves the determination of the gravitational potential using the brightness profile of the galaxy, the construction of an orbit library, the projection of the orbits on the plane of the sky, and then the superposition of the orbits, in order to reproduce the observed kinematics. Usually, the Schwarzschild method is applied assuming that the galaxy is axisymmetric or triaxial. The Jeans method usually assumes axisymmetry and involves the solution of the Jeans equations, in order to obtain the gravitational potential and the kinematics around the SBH. Although our method is not as detailed as these two, it has certain advantages. The Schwarzschild method, for example, is frequently affected by degeneracies and difficulties in the convergence. Such problems are rarely observed in the model of a thin eccentric disk, due to its simplicity. The Jeans method often assumes axisymmetry, which is not required by the thin disk method. We have already used the implementation of the Jeans method developed by \citet{cap08} to try to simulate the kinematic maps of the data cube of M104 \citep{men12}. However, we were not able to reproduce certain aspects of these maps, like the displacement of $\sigma_*(peak)$ from the $V_*$ map center. This can be explained by the fact that the implementation of the Jeans method we used assumes axisymmetry, which is not the most appropriate approximation to be used for this galaxy. As we verified in this work, the assumption of a rotating eccentric disk is more adequate. In summary, the disadvantages of our model of a thin excentric disk are: it, obviously, assumes the existence of a thin rotating disk, which is not necessarily the case; it is less detailed than the Schwarzschild and the Jeans methods. On the other hand, its advantages are: due to its simplicity, it is usually not affected by degeneracies and problems with convergence; it does not require axisymmetry. The fact that the mass of the SBH we obtained is compatible, at $1\sigma$ or $2\sigma$ levels, with the values found in previous studies using very detailed dynamical models indicates that the model of a thin eccentric disk can be used to describe, with good precision, the stellar kinematics around the SBH in M104. 

The presence of an eccentric disk around a SBH of a galaxy is not actually new, as this structure was already detected in the nuclear region of M31, the Andromeda galaxy. The eccentric disk in this object is responsible for the existence of a double nucleus \citep{lau93}. The two components are called P1 (the brightest one) and P2 (located, approximatelly, at the center of the stellar bulge). P1 coincides with the apocenter of the stellar disk, while P2 is located in the vicinity of the SBH. \citet{ben05} verified that the SBH is actually located in a structure embedded in P2, called P3, which is probably a cluster of A type stars. \citet{lau12} confirmed that P3 is a cluster of blue stars around the central black hole. Different studies detected an asymmetry in the stellar rotation curve of the nuclear region of M31 and also a displacement between the kinematic center and the peak of the stellar velocity dispersion \citep{bac94,bac01,ben05}. This last feature is located in an area close to the pericenter of the disk, which shows a lower surface brightness than the area corresponding to the apocenter (P1). All these photometric and kinematic characteristics were successfully reproduced with models involving a rotating eccentric stellar disk around the SBH \citep{tre95,sal04,pei03}. In the case of M104, our dynamical modeling revealed that the eccentricity of the stellar disk is smaller than the values obtained for M31 (which can be higher than 0.3 for certain parts of the disk, as revealed by the studies mentioned before). Nevertheless, we can identify many similarities in the stellar kinematics and in the photometry of these two objects. The kinematic similarities are the asymmetry in the rotation curve and the displacement between the kinematic center and the peak of the stellar velocity dispersion. In the case of the photometry, the similarities are more difficult to detect. A careful analysis of the \textit{HST} images in Figure~\ref{fig4} reveals a slight decrease in the surface brightness of the disk in an area at a distance of $\sim 0\arcsec\!\!.25$ eastward from the nucleus. This region corresponds approximatelly to the position of $\sigma_*(peak)$ and, according to our dynamical modeling, it also coincides with the pericenter of the disk. On the other hand, the surface brightness of the disk is higher in the diametrically opposed area, which, according to our dynamical modeling, coincides with the apocenter of the disk. Since the purpose of this work is not to analyze in detail the photometry of the nuclear region of M104, we will not give further detail about this topic. However, the discussion above reveals that the photometry and the stellar kinematics in the nuclear regions of M104 and M31 (which shows a well studied rotating eccentric stellar disk) have certain similarities, although some of them are subtle. This indicates that the model of an eccentric stellar disk is actually a very natural choice to reproduce the stellar kinematics in the nuclear region of M104.

\section{Summary and conclusions}

We analyzed a data cube of the nuclear region of M104, the Sombrero galaxy, obtained with NIFS on the Gemini-north telescope. After a subtraction of the stellar continuum, we detected the H$_2 \lambda 21218$ emission line in the spectra of the data cube. The image, and also the corresponding RGB image, of this line indicate that the molecular gas is disposed along a rotating torus/disk structure, aproximately co-planar with the inner stellar disk around the nucleus of M104. This molecular structure is apparently coplanar with another structure, made of dust, detected by us in \citet{men13}, which we identified as an obscuring torus/disk. Therefore, we conclude that these two findings are probably associated with the same obscuring torus/disk.

Using the pPXF method in the data cube of M104, we obtained maps of $V_*$, $\sigma_*$, $h_3$, and $h_4$. We detected a rotating stellar disk around the nucleus. We also observed that the $\sigma_*$ values along the stellar disk are considerably smaller than the values at the surrounding areas. This indicates that the stellar kinematics in the nuclear region of M104 can be described by the superposition of a ``cold'' rotating disk (with smaller $\sigma_*$ values) and a ``hot'' bulge (with larger $\sigma_*$ values). Assuming the existence of a thin eccentric disk, we applied a dynamical modeling, in order to reproduce the $V_*$, $\sigma_*$, and $h_3$ maps and also obtain an estimate of several kinematic parameters. The value of the mass of the SBH we obtained ($M_{\bullet} = 9.0 \pm 2.0 \times 10^8 M_{\sun}$) is compatible, at $1\sigma$ or $2\sigma$ levels, with the estimates found by many previous studies \citep{kor88,ems94,kor96,mag98,jar11}. Our modeling reproduced the main properties of the $V_*$ and $\sigma_*$ maps, specially within a distance of $0\arcsec\!\!.2$ from the kinematic axis, and also the general behavior of the $h_3$ map (although the discrepancies, in this case, are considerably larger). For distances of $0\arcsec\!\!.2 - 0\arcsec\!\!.4$  from the kinematic axis, the discrepancies between the simulated and observed kinematic maps increase substantially, which is probably a consequence of the limitation of a thin eccentric disk model. Nevertheless, considering all the results and also the compatibility between the value of $M_{\bullet}$ obtained by us and by many previous studies, we conclude that the model of thin eccentric disk is a reasonably good description for the stellar kinematics around the SBH in M104.    

The model of an eccentric disk around the SBH of a galaxy was already applied to reproduce the stellar kinematics around the nucleus of M31. Considering that, and also the results obtained in this work, we believe that this model may also be used to reproduce the stellar kinematics in the nuclear region of other galaxies and obtain reliable estimates for the masses of their central SBHs.

\acknowledgments

Based on observations obtained at the Gemini Observatory (processed using the Gemini IRAF package), which is operated by the Association of Universities for Research in Astronomy, Inc., under a cooperative agreement with the NSF on behalf of the Gemini partnership: the National Science Foundation (United States), the National Research Council (Canada), CONICYT (Chile), the Australian Research Council (Australia), Minist\'{e}rio da Ci\^{e}ncia, Tecnologia e Inova\c{c}\~{a}o (Brazil) and Ministerio de Ciencia, Tecnolog\'{i}a e Innovaci\'{o}n Productiva (Argentina). We thank FAPESP for support under grants 2012/02268-8 (RBM) and 2011/51680-6 (JES) and also an anonymous referee for valuable comments about this article.

{\it Facilities:} \facility{Gemini:Gillett(NIFS)}.

\end{document}